%% file: kms-network.tex
\documentclass[pre,superscriptaddress,twocolumn]{revtex4-1}

\usepackage{adjustbox}
\usepackage{amssymb}
\usepackage{bm}
\usepackage{makecell}
\usepackage{microtype}
\usepackage{multirow}
\usepackage{paralist}
\usepackage{rotating}
\usepackage{indentfirst}
\usepackage{float}
\usepackage{mdframed}
\usepackage{hyperref}
\usepackage[all]{xy}
\usepackage[table]{xcolor}
\usepackage{dcolumn}
\usepackage{amsmath, amsthm, amssymb,amscd, mathrsfs, amsfonts, mathtools}
\usepackage{tikz}
\usepackage{diagbox}
\usepackage{cleveref}
\usetikzlibrary{matrix,arrows,decorations.pathmorphing}
\usetikzlibrary{backgrounds}
\usetikzlibrary{shapes.geometric}
\usepackage{circuitikz}
\usepackage{tcolorbox}
\usepackage{blkarray}
\usepackage{subcaption}
\usepackage{multirow}

\usepackage{xspace}
\usepackage{ifthen}

\newcommand{\ie}{i.e.\@\xspace}

\newcommand{\wrt}{w.r.t.\@\xspace}
\newcommand{\cels}{{\em C. elegans}\@\xspace}

\def\rTo{\longrightarrow}
\def\rto{\rightarrow}
\def\lto{\leftarrow}
\newcommand{\mto}{\mapsto}

\def\<{\langle}
\def\>{\rangle}
\def\path{\rightsquigarrow}

\newcommand{\inner}[2]{\langle{#1},{#2}\rangle} 

\newcounter{para}

\def\a{\alpha}
\def\b{\beta}

\def\o{\omega}
\def\g{\gamma}
\def\t{\theta}
\def\d{\delta}
\def\l{\lambda}

\def\s{\sigma}

\let\cal\mathcal 
\let\bb\mathbb 

\def\OG{{\cal O}_G} 

\def\PSI{\operatorname{\bf \Psi}}

\def\Tr{\operatorname{Tr}}

\def\pre{\operatorname{\tt pre}}
\def\post{\operatorname{\tt post}}
\def\xx{\operatorname{{\bf x}}}
\def\yy{\operatorname{{\bf y}}}
\def\pp{\operatorname{\bf p}}

\def\bc{4.294}
\def\bs{10.74}
\def\bgel{15.035}

\tikzset{every loop/.style={looseness=10, in=-130, min distance=1mm}} 
\tikzset{1simpl/.style={->,>=stealth,thick}}
\tikzset{vertex/.style = {circle, fill=red!20, inner sep=8}}
\tikzset{edgeto/.style={line width=1.2, draw, arrows={-latex}, color=black!70 }
}

\allowdisplaybreaks

\begin{document}
	
	\title{Kubo-Martin-Schwinger states of Path-structured Flow in Directed Brain Synaptic Networks}

	\begin{abstract}
	   The brain's synaptic network, characterized by parallel connections and feedback loops, drives interaction pathways between neurons through a large system with infinitely many degrees of freedom. This system is best modeled by the graph C*-algebra of the underlying directed graph, the Toeplitz-Cuntz-Krieger (TCK) algebra, which captures the diversity of path-structured flow connectivity. Equipped with the gauge action, the TCK algebra defines an {\em algebraic quantum system}, and here we demonstrate that its thermodynamic properties provide a natural framework for describing the dynamic mappings of potential flow pathways within the network. Specifically, the KMS states of this system represent the stationary distributions of a non-Markovian stochastic process with memory decay, capturing how influence propagates along exponentially weighted paths, and yield global statistical measures of neuronal interactions. Applied to the {\em C. elegans} synaptic network, our framework reveals that neurolocomotor neurons emerge as the primary hubs of incoming path-structured flow at inverse temperatures where the entropy of KMS states peaks. This finding aligns with experimental evidence of the foundational role of locomotion in {\em C. elegans} behavior, suggesting that functional centrality may arise from the topological embedding of neurons rather than solely from local physiological properties. Our results highlight the potential of algebraic quantum methods and graph algebras to uncover patterns of functional organization in complex systems and neuroscience.

	\end{abstract}
	\author{Elka\"ioum M.~Moutuou}
	\email[Corresponding author: ]{elkaioum.moutuou@concordia.ca}
	\affiliation{Department of Electrical and Computer Engineering, Concordia University, Montreal, QC, H3G 1M8}
	\affiliation{Centre de Recherche de l’Institut Universitaire de G\'eriatrie de Montr\'eal (CRIUGM), Montr\'eal, QC, Canada}
	\author{Habib Benali}
	\email[Corresponding author: ]{habib.benali@concordia.ca}
	\affiliation{Department of Electrical and Computer Engineering, Concordia University, Montreal, QC, H3G 1M8}
	\affiliation{Centre de Recherche de l’Institut Universitaire de G\'eriatrie de Montr\'eal (CRIUGM), Montr\'eal, QC, Canada}
	\affiliation{Inserm U1146, Paris, France}
	\maketitle
	
Understanding the dynamics of path-structured flow patterns in complex systems is a fundamental challenge in network science and is crucial in neuroscience for uncovering the mechanistic basis of brain function~\cite{emmons2012mood,Honey2010,Randi2021}. While various approaches have been proposed~\cite{Stojmirovic2012,Harush2017,Ghavasieh2024,Bettencourt2008},
systematic methodologies for accurately describing states of path-structured propagation within directed networks remain lacking. Here, using the example of the \cels synaptic network as a finite directed graph $G$ with parallel edges and self-loops, we study 
its path-structured flow connectivity as a system of infinitely many interacting subsystems.

Due to multiple chemical synapses, reciprocal electrical channels via gap-junctions, and self-synapses, 
$G$ contains directed cycles and recurrent patterns, resulting in interactions with infinitely many degrees of freedom. Algebraic quantum mechanics (AQM)~\cite{hugenholtz1972} and graph C*-algebras~\cite{raeburn2005graph} offer sophisticated formalism to describe such a system. The main idea behind AQM methods
~\cite{haag1967quantum,hugenholtz1972} is that physical states are represented by {\em states} of a C*-algebra $\cal O$, unlike standard quantum mechanics, which uses a Hilbert space of state vectors and typically deals with systems with finite degrees of freedom. 
Observables are self-adjoint elements in $\cal O$, and time-evolution is represented by a $C^*$--dynamics $\a_t$ on $\cal O$. The pair $(\cal O, \a)$ is referred to as a {\em C*--dynamical system} or an {\em algebraic quantum system} (see Appendix~\ref{Appendix:AQM}). 

Given a state $\o$ and an observable $U\in \cal O$, the number $\o(U)$ is the expectation value of $U$ in state $\o$. The observable $U$ is {\em analytic} \wrt $\a$ if $t\mto \a_t(U)$ extends to an analytic function $\bb C \rTo \cal O, \ z\mto \a_z(U)$, where $z=t+ i\b$. Equilibrium states of $(\cal O,\a)$ at inverse temperature $\b$ are states $\o$ satisfying the Kubo-Martin-Schwinger (KMS) condition~\cite{haag1967quantum,bratteli1982equilibrium,bost-connes1995}:
\begin{eqnarray}\label{eq:KMS}
	\o(U\a_{i\b}(V))= \o(VU),
\end{eqnarray}
for all analytic elements $V,U$. 
Such $\o$ is called a {\em KMS state}. This formalism is an algebraic extension~\cite{hugenholtz1972} of classical quantum statistical mechanics~\cite{Hall2013}, where observables are represented by self-adjoint matrices, time-evolution is generated by a Hamiltonian $H$ via $\a_t(U)=e^{itH}Ue^{-itH}$, and equilibrium states are given by the Gibbs state $\o_\b(U) =\Tr(\rho U) $, with $\rho = e^{-\b H } / \Tr(e^{-\b H})$.

Here, the use of this formalism is motivated by its ability to rigorously describe the dynamics of path-structured flow, though our framework does not involve traditional quantum phenomena. $\cal O$ is the {\em Toeplitz-Cuntz-Krieger (TCK) algebra}~\cite{raeburn2005graph} of $G$, capturing all possible flow pathways through operators on an infinite-dimensional Hilbert space. Equipped with its natural dynamics, the KMS states of the resulting algebraic quantum system are generated by probability distributions, interpreted as stationary states of a generalized non-Markovian stochastic process with exponentially decaying memory (see Appendix~\ref{Appendix:process}), that describe the interaction profiles of all neurons. These KMS states provide a rigorous foundation for uncovering the mechanisms through which network structure drives the dynamic states of functional interactions. While this work focuses on establishing and illustrating the conceptual foundations of our AQM-based framework, its functional implications and detailed applications are presented in a companion study~\cite{Moutuou2024b}.

\subsection{Directed synaptic networks} 
We  consider a complex system represented by a directed multigraph $G$ with $N$ nodes $i, j$, etc., directed edges $a, b$, etc., and  adjacency matrix $A$, where $A_{ij}$ is the number of directed edges from node $j$ to node $i$. Without loss of generality, $G$ represents a directed network whose nodes are neurons and edges are chemical and electrical synapses~\cite{white1986structure,Varshney2011}; thus, $A_{ij}$ is the number of synapses from neuron $j$ to $i$. We write $\pre(a)=j$ and $\post(a)=i$ if $a$ is a synapse with pre-synaptic neuron $j$ and post-synaptic neuron $i$. A sequence $\g=a_n\cdots a_1$ of synapses $a_1, ..., a_n$ (with possible recurrences) such that $\pre(a_{l+1})=\post(a_l)$ is a {\em synaptic walk} or {\em paths} of length $|\g|=n$, with $\pre(\g)=\pre(a_1)$ and $\post(\g)=\post(a_n)$. Paths concatenate if $\pre(\g) = \post(\t)$, in which case $\g\t$ is also a path.

\subsubsection{Algebraic operations}  
\begin{figure}[!h]
	\includegraphics[width=.45\textwidth]{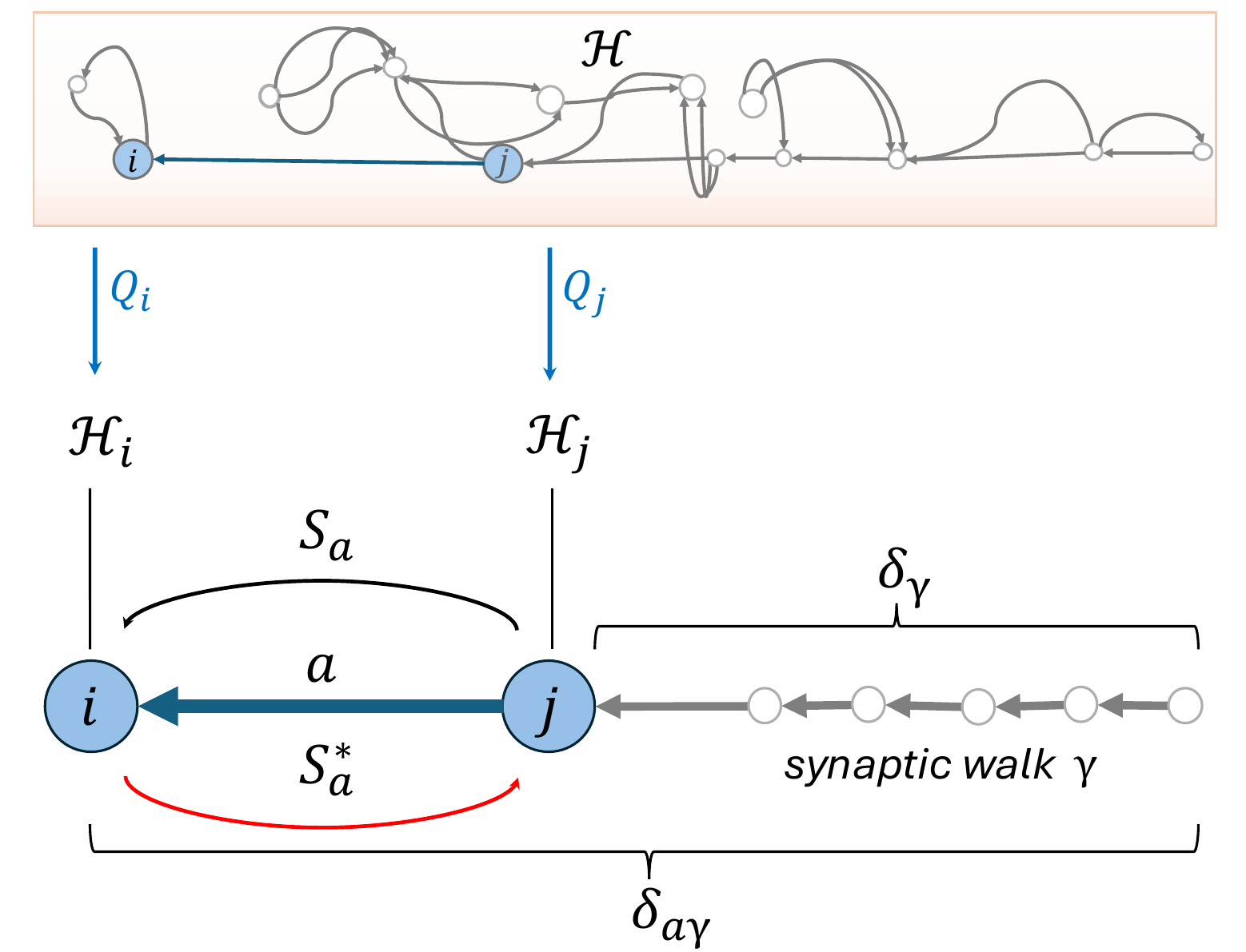}
	\caption{The Hilbert space $\cal H$ is spanned by scaled paths $\d_\g$. The projection $Q_j$ selects the subspace $\cal H_j$ of paths ending at neuron $j$. The operator $S_a$ extends a path $\d_\g$ ending at neuron $j$ to $\d_{a\g}$ reaching neuron $i$, while $S_a^*$ extracts the portion of flow flow to $i$ that passes through $j$. }
	\label{fig:operators}
\end{figure}

We define algebraic operations on paths to describe concatenation and scaling. Let $\cal H$ be the infinite-dimensional Hilbert space 
of complex-valued sequences $h=(\xi_\g)_\g$ indexed by synaptic walks $\g$, such that $\sum |\xi_\g|^2<\infty$. Elements of $\cal H$ are expressed as formal sums  $h=\sum_\g \xi_\g\d_\g$, where $\d_\g$ is the Dirac measure on path $\g$, and $\xi_\g$ are path scaling factors. To neuron $i$, we associate the projection $Q_i$ onto the subspace $\cal H_i$ generated by all $\d_\g$ with $\post(\g)=i$. And to synapse $a:j\rto i$, we associate a partial isometry $S_a$ from $\cal H_j$ to $\cal H_i$, defined by $S_a(\d_\g) = \d_{a\g}$ if $\post(\g) = \pre(a)$, and $0$ otherwise. The adjoint is given by $S_a^*(\d_{a\g}) = \d_\g$ (see Appendix~\ref{Appendix:AQM}). These representations are illustrated in Fig.~\ref{fig:operators}).

The {\em TCK algebra}~\cite{raeburn2005graph} is the universal C*-algebra $\cal O$ 
generated by the family $\{Q_i, S_a\}$ subject to 
\begin{eqnarray}\label{eq:TCK}
	S_{a}^*S_a = Q_j \ {\rm for \ } \pre(a) = j, \ \sum_{\post(a) = i}S_aS_a^* \le Q_i.
\end{eqnarray}
Namely, $\cal O$ is spanned by products $S_{\g}S_{\t}^*$ with $\pre(\g)=\pre(\t)$, where for  $\g=a_n\cdots a_1$, we set $S_{\g}=S_{a_n}\cdots S_{a_1}$. 

\subsection{The graph algebraic quantum system}  
The algebra $\cal O$ carries a C*-dynamics $\a_t$, obtained from the {\em gauge action}:
\begin{eqnarray}\label{eq:dynamics}
\a_t(S_\g S_\t^*) = e^{it(|\g|-|\t|)}S_\g S_\t^*,
\end{eqnarray}
defining the algebraic quantum system $(\cal O,\a)$ (see Appendix). Any state $\o$ on $(\cal O,\a)$ defines a probability distribution $\xx^\o_i$ over neurons via $\xx^\o_i = \o(Q_i)$, since $\o(Q_i) \ge 0$ and $\o(\sum Q_i) = 1$, as $\sum_iQ_i$ is the unit in $\cal O$ (see Appendix~\ref{Appendix:AQM}). Now by Eq.~\ref{eq:KMS} and~\eqref{eq:dynamics}, a state $\o$ of $(\cal O,\a)$ is KMS at inverse temperature $\b$ iff  
\[
\begin{aligned}
	\o(S_\g S_\t^*) & = \o(S_\t^*\a_{i\b}(S_\g)) \\
	& =  e^{-\b|\g|}\o(S_\t^*S_\g)   \\
	& =  e^{-\b|\g|}\o(S_\g \a_{i\b}(S_\t^*))\\
	& =  e^{-\b(|\g| - |\t|)} \o(S_\g S_\t^*).
\end{aligned}
\]
Hence, $\o(S_\g S_\g^*) = e^{-\b |\g|}\o(Q_{\pre(\g)})$ as a consequence of the second line of this equation, while the last line yields  $\o(S_\g S_\t^*)  = 0$ if $|\g| \neq |\t|$. In particular, for a synapse $a$, the expected value of $S_aS_a^*$ in a KMS state $\o$ satisfies
\begin{eqnarray}\label{eq:KMS-graph2}
	e^{\b}\o(S_aS_a^*) =  \xx^\o_{\pre(a)},
\end{eqnarray}
which, thanks to Eq.~\eqref{eq:TCK} combined with Eq.~\eqref{eq:KMS-graph2}, implies $\sum_{\post(a)=i}\xx^\o_{\pre(a)} \le e^\b \xx^\o_i$. Using the adjacency matrix $A$, $\sum_j A_{ij}\xx^\o_j \le e^\b \xx^\o_i$; \ie, $A\xx^\o \le e^{\b}\xx^\o$.  If $\b > \log r$, where $r$ is the spectral radius of $A$, $\xx^\o$ is not an eigenvector of $A$ and $1 - e^{-\b}A$ is non-singular and the components of $\Psi^\o = (1 - e^{-\b}A)\xx^\o$ are non-negative, so: 
\begin{eqnarray}\label{eq:xi-psi}
	\xx^\o = (1-e^{-\b}A)^{-1}\Psi^\o.
\end{eqnarray}
To characterize the vectors $\Psi$ associated with KMS states as in Eq.~\ref{eq:xi-psi}, define the {\em partition vector} $Z^\b$, where $Z^\b_j  =  \sum_{\pre(\g) = j} e^{-\b |\g|}$, which counts all paths starting at $j$ weighted by the propagation scale $e^{-\b}$. Since $\b > \log r$, the expansion$(1-e^{-\b}A)^{-1}=\sum_ne^{-\b n}A^n$ converges (see Appendix~\ref{Appendix:AQM}), and using the fact that $A^n_{ij}$ counts all walks from $j$ to $i$, we have
\[
	Z_j^\b  =  \sum_n\sum_ie^{-\b n} A^n_{ij} = \sum_i(1-e^{-\b}A)_{ij}^{-1}.
\]
Then, $\Psi^\o$ satisfies: 
\begin{eqnarray}\label{eq:psi-state}
	\langle \Psi, Z^\b \rangle = 1.
\end{eqnarray}   
Conversely, if a non-negative vector $\Psi$ satisfies Eq.~\eqref{eq:psi-state}, then $\xx = (1-e^{-\b}A)^{-1}\Psi$
 is a probability distribution and defines a KMS state $\o^{\Psi}$ via $\o^{\Psi}(S_\g S_\t^*) = e^{-\b |\g|}\xx_{\pre(\g)}$ if $\g=\t$, and $0$ otherwise~\cite{huef2012KMS}. Thus, for $\b > \log r$, the KMS state space $\Omega_\b$ is determined by Eq.~\eqref{eq:psi-state} and the values $\o(Q_i)$ (see Appendix~\ref{Appendix:AQM} for details). 

\subsubsection{Neuronal interactions}
For $\b > \log r$, the KMS state space $\Omega_\b$ is a convex set~\cite{hugenholtz1972}, and its extreme points are the {\em pure KMS states}. These are computed explicitly: for fixed neuron $j$, $\Psi^{j|\b} = \left(0, ...,0, (Z_j^\b)^{-1},0, ...,0\right)$ solves Eq.~\eqref{eq:psi-state}, and defines the pure KMS state $\xx^{j|\b} = (1-e^{-\b}A)^{-1}\Psi^{j|\b}$. Thus, there is a 'critical' inverse temperature $\b_c = \log r$ above which $\Omega_\b$ is the $(N-1)$--dimensional convex hull of the pure states $\xx^{j|\b}$. The $i$-component of $\xx^{j|\b}$ has the explicit form:
\begin{eqnarray}
		\xx^{j|\b}_i   =  \frac{1}{Z_j^\b} \sum_ne^{-\b n}A^n_{ij} 
		  =  \frac{1}{Z_j^\b} \sum_{\substack{\post(\g)=i \\ \pre(\g)=j}}e^{-\b |\g|},
\end{eqnarray}
giving the probability of interaction from neuron $j$ to neuron $i$ at inverse temperature $\b$.

Thus, $\xx^{j|\b}$ is the outgoing path-structured {\em interaction profile} of $j$ at a fixed $\b$. Varying $\b$ from larger $\b$ toward the critical value $\b_c$ draws the dynamic mapping of all potential interactions originting from $j$. Specifically, for sufficiently large $\b$, $\xx^{j|\b}\sim (0, ...,0,1,0,...,0)$, where $1$ is at the $j$-th component, indicating purely local self-interaction. As $\b$ approaches $\b_c$, the profile spreads, gaining more and more nonzero components, while the self-interaction weight $\xx^{j|\b}_j$ decreases.

\subsubsection{Mixed states and dynamic flow patterns}
 Mixed KMS states are obtained as statistical superpositions of the $\xx^{j|\b}$ via
\begin{eqnarray}\label{eq:mixed-state}
	\xx = \sum_j\pp_j\xx^{j|\b},
\end{eqnarray}
where $\pp = (\pp_j)_j$ is a probability distribution over neurons, representing relative quantities such as gene expression, external stimuli, or topological measures. A mixed KMS state thus represents the expected interaction profile of the network under a given distribution $\pp$, capturing how relative concentrations influence the dynamics of neural interactions. 

For instance, let $C$ be a subset of $k$ neurons and define $\pp$ as the uniform distribution on $C$ (\ie, $\pp_j = 1/k$ if $j\in C$, and $0$ otherwise). Then $\langle C\rangle^\b = \frac{1}{k}\sum_C\xx^{j|\b}$ is the average outgoing interaction profile of $C$ at inverse temperature $\b$. Varying $\b$ and visualizing $\langle C\rangle^\b$ as weighted directed networks reveal the dynamic flow patterns emerging from $C$.

\subsubsection{Illustration of KMS matrices on a synthetic network} 
\begin{figure}[!h]
	\centering
	\begin{tabular}{c}
		(a) \\
		\input{SM/figures/toy} 
		\\
		(b) \\
		\includegraphics[width=.4\textwidth]{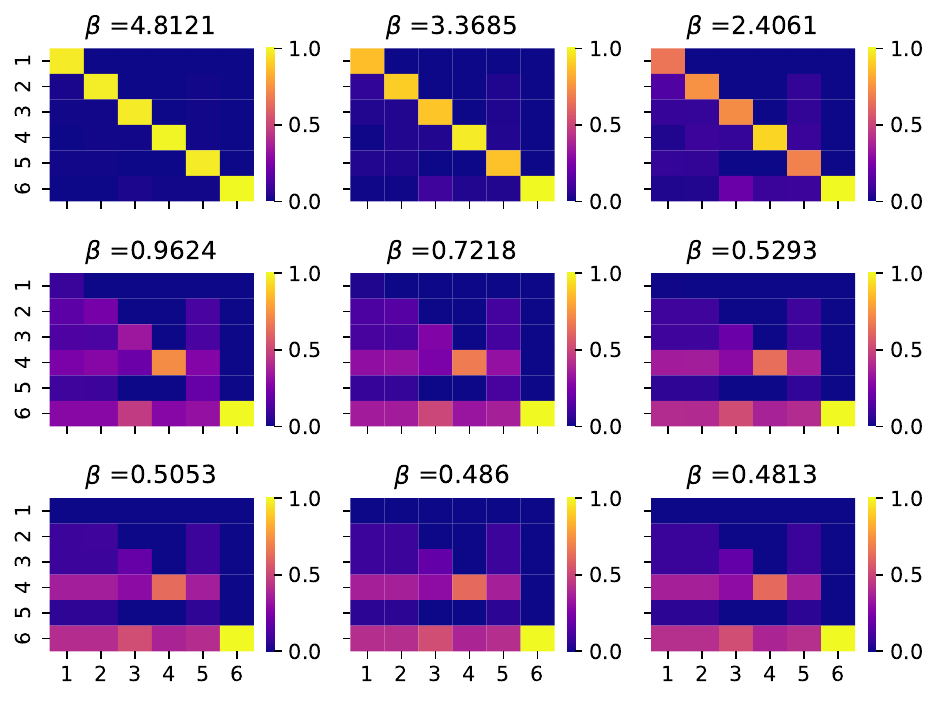}
		
	\end{tabular}
   \caption{{\bf Schematic illustration of a directed network and its KMS matrices.} (a) A network represented by a directed multigraph $G$ with $N=6$ nodes and 18 edges, including multiple parallel edges and self-loops. (b) Connectivity matrices defined by the KMS state matrices of $G$. The column representing the pure KMS state $\xx^{6|\b}$ is the unit vector $(0,0,0,0,0,1)$ for all values of $\b$, as node $6$ has no outgoing edges} 
	\label{fig:toy}
\end{figure}

By Eq.~\eqref{eq:mixed-state}, mixed states depend on the column-stochastic matrix $[\xx^{\bullet|\b}]$ (the KMS matrix) whose columns are the vectors $\xx^{j|\b}$. These matrices define weighted directed networks representing information flow at inverse temperature $\b$. To illustrate how these flow networks change with respect to $\b$, we  consider the synthetic directed multigraph $G$ illustrated in Fig.~\ref{fig:toy}a. Its critical inverse temperature is $\b_c= \log r = 0.4812$. 

The KMS state matrices $[\xx^{\bullet|\b}]$ of $G$ are computed and presented in Fig.~\ref{fig:toy}b at different $\b$-values. At large $\b$ values ($\b \le 4.8$), $[\xx^{\bullet|\b}]$ is the identity matrix $I_6$, reflecting the fact that at lower temperature, all pathways are 'frozen' and no information can flow out of the nodes, therefore the pure KMS states at these temperatures are all unit vectors. As $\b$ decreases, the diagonal values decreases and the matrices become less sparse. Indeed, as temperature increases (lower values of $\b$), more and more flow pathways are 'active', allowing more long range interactions between nodes. Notice that at all $\b$ values, the column corresponding to the pure KMS state $\xx^{6|\b}$ remains the unit vector with $1$ at the $6^{th}$ position; this reflects the fact that node $6$ has no out-going connections (Fig.~\ref{fig:toy})

\subsubsection{KMS states entropy quantifies flow selectivity}
Eq.~\eqref{eq:mixed-state} naturally defines the {\em entropy of the system in equilibrium} at inverse temperature $\b$ via Shannon entropy:
\begin{eqnarray}
	\cal S(\xx^{\bullet|\b},\pp) = \cal S(\xx) = -\sum_i\yy_i^\b\log \yy_i^\b,
\end{eqnarray}
where $\yy_i^\b= \sum_j\pp_j\xx^{j|\b}_i$. This entropy quantifies uncertainty in neuronal interactions at scale $\b$ under the distribution $\pp$. High entropy reflects "non-selective", diffusive interactions; low entropy indicates stronger selectivity in outgoing pathways. 

In particular, the entropy of $\langle C\rangle^\b$, denoted $\cal S(\langle C\rangle, \b)$, quantifies the average interaction selectivity of the group $C$. When$C = \{j\}$, we recover the entropy of the single neuron's profile $\cal S(\xx^{j|\b}) =  \cal S(\langle \{j\}\rangle, \b) =-\sum_i\xx^{j|\b}_i\log \xx^{j|\b}_i$, which quantifies how selectively neuron $j$ distributes its outgoing flow. 

For a fixed topology, this selectivity depends only on $\b$ and $\pp$. Indeed, $\cal S(\xx^{\bullet|\b},\pp)$ is maximal if $\yy^{\b}_i = 1/N$ for all $i$, yielding $\cal S(\xx^{\bullet|\b},\pp) = \log N$. As shown in Fig.~\ref{fig:entropy}a, this occurs when $\pp$ is the uniform distribution and at sufficiently low temperatures (larg values of $\b$), where the profiles $\xx^{j|\b}$ approximate the unit vectors corresponding to trivial self-interactions.

\subsubsection{Phase transitions and Symmetry breaking}
For large $\b$, the system has different states generated by $\xx^{j|\b}$. In contrast, $(\cal O, \a)$ has no KMS states at $\b< \b_c$, and its KMS state at $\b_c$ is obtained as limit of pure states as $\b \to \b_c$~\cite{huef2012KMS}. Specifically, the profiles $\xx^{j|\b}$ converge to a common state as $\b$ approaches $\b_c$ where disorder is predominant, and differentiate as temperature decreases to allow for more order. This behaviour reflects a symmetry breaking in path-structured flow: at high temperature near $\sim T_c=1/\b_c$, pure KMS states yield symmetric flow networks, while low temperatures lead to increasingly asymmetric interaction profiles. To quantify this asymmetry, we consider the {\em transition probability} introduced by Uhlmann~\cite{uhlmann1976} and later used by Jozsa~\cite{Jozsa1994} to define 'fidelity' between quantum states. In our context, the transition between two profiles $\xx^{j|\b}$ and $\xx^{j'|\b}$ is given by:
\begin{eqnarray}
	P(j,j',\b) = \left( \sum_i \left(\xx^{j|\b}_i\xx^{j'|\b}_i \right)^{1/2}\right)^2,
\end{eqnarray}
This quantity $P(j,j',\b)\in [0,1]$ measures the similarity between flow patterns driven by neurons $j$ or $j'$. A higher value indicates that the corresponding flow networks are nearly indistinguishable, while lower value signals significant structural differences. Statistically, $P(j,j',\b)$ is the probability of correctly identifying the system's state at inverse temperature $\b$, given that it is in one of the pure states $\xx^{j|\b}$ or $\xx^{j'|\b}$. In physical terms, it quantifies the symmetry between flow profiles of two neurons.

 \subsection{Application} 
Let $G$ represent the \cels network of $N=280$ non-pharyngeal neurons and their 12071 synaptic connections reconstructed from~\cite{white1986structure,Varshney2011,cook2019connectome}. In this example, $\b_c = \bc$.

 \begin{figure}[!h]
 	\includegraphics[width=.45\textwidth]{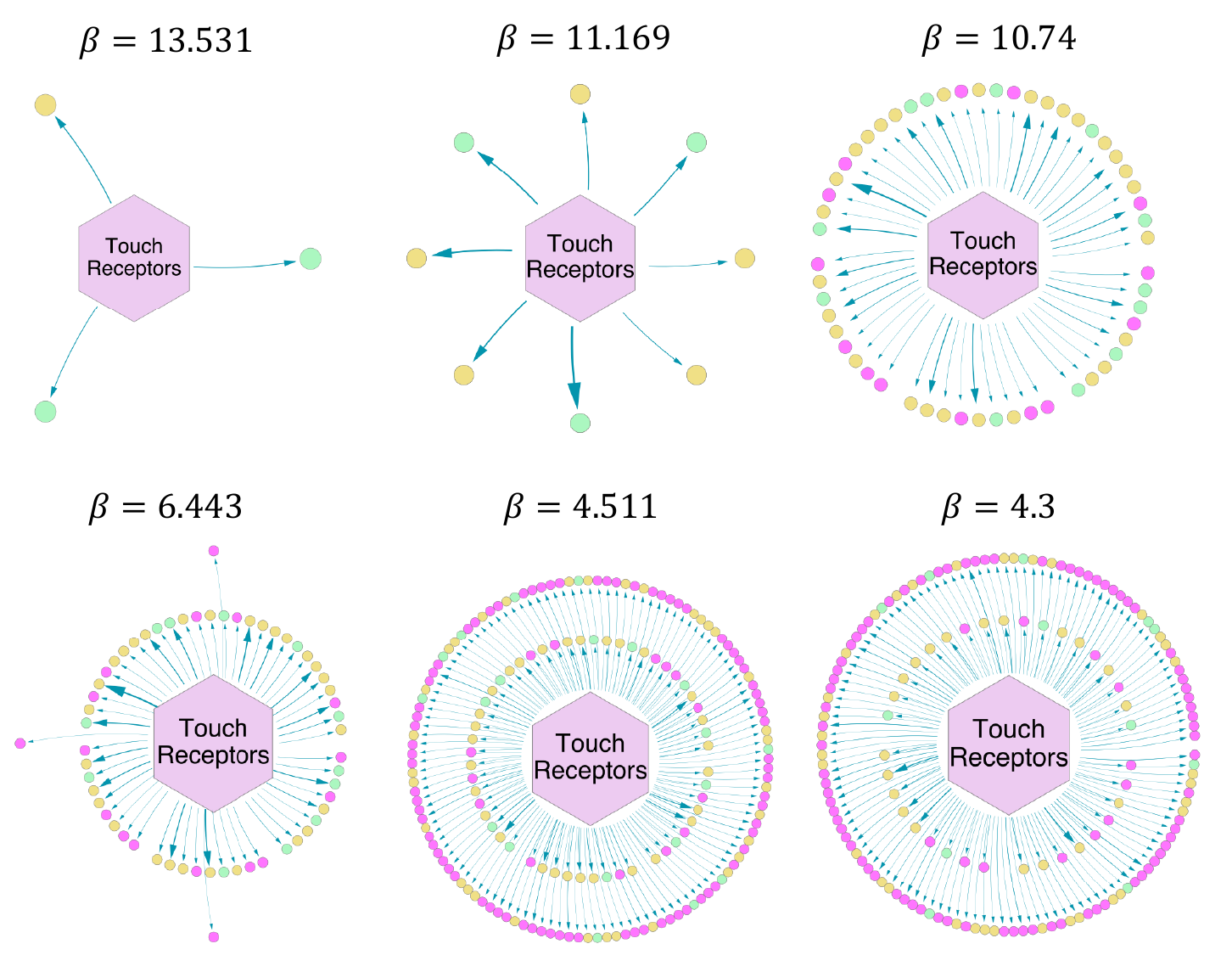}
 	\caption{Average path-structured flow dynamics from touch receptor neurons. At low temperatures (high $\b$), $C$ interacts only with direct neighbors; higher temperatures enable long-range interactions, mainly targeting motor neurons. Colors: green (sensory), yellow (inter), purple (motor). Self-interactions omitted.}
 	\label{fig:patterns}
 \end{figure}

\subsubsection{Dynamic flow patterns}

 \begin{figure*}[!th]
 	\includegraphics[width=\textwidth]{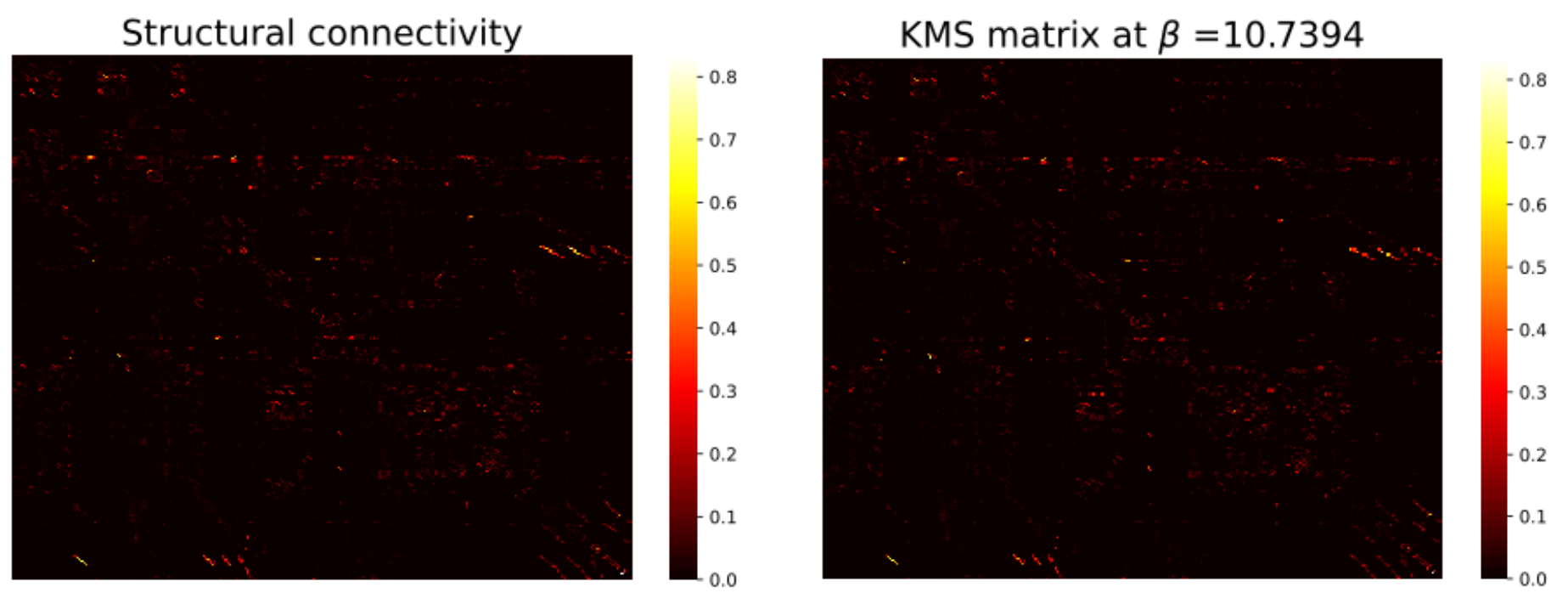}
 	\caption{Comparison between weighted structural connectivity of \cels directed synaptic network and its KMS matrix at inverse temperature $\b=10.7394$ (about 2.5 times the critical value $\b_c = 4.295$).}
 	\label{fig:kms-matrix-cels}
 \end{figure*}	
 
To illustrate mixed states $\langle C\rangle^\b$, we examine the touch receptor subset $C = \{ALML/R, PLML/R, AVM,PVM\}$~\cite{kaplan1993,Goodman2006}(Fig.~\ref{fig:patterns}). For $\b \ge \bgel$, interactions are limited to trivial self-loops. Around $\b \approx 3\b_c$ ($\b \ge 13.531$), connections emerge to immediate neighbors. At $\b = 2.5\b_c=\bs$, the network mirrors the density of direct physical connections. Below this threshold, longer-range connections appear as local interactions diminish.

\subsubsection{Mixed KMS states reveal centrality of locomotion in C. elegans}
Consider the adjacency matrix $C$ of the weighted directed graph representing \cels connectome with synapse counts between neurons replaced by their relative weights with respect to outgoing synapses; \ie, the weight of a connection from $j$ to $i$ is the ratio of the number of synapses from $j$ to $i$  out of the number of all outgoing synapses from $j$. We represent $C$ as the {\em structural connectivity} in Fig.~\ref{fig:kms-matrix-cels}. As shown in Fig.~\ref{fig:kms-matrix-cels}, this matrix is approximated by the KMS state matrix $[\xx^{\bullet|\b}]$ at inverse temperature $\b = 10.7394$. This means that at this $\b$ value, information flow is localized so that the only possible interactions are between direct neighbors.

\begin{figure}[!h]
 	\includegraphics[width=.45\textwidth]{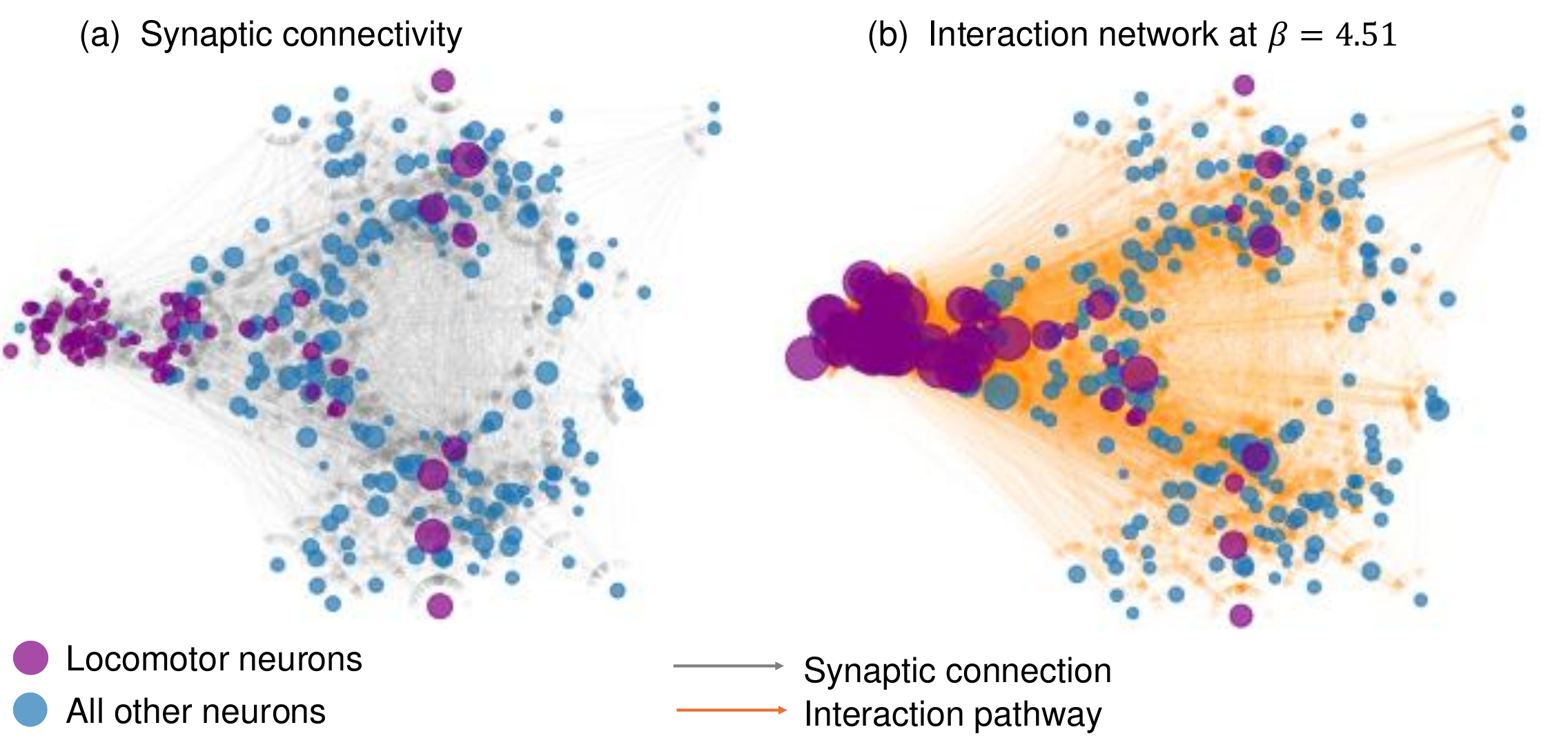}
 	\caption{Binarized directed networks of (a) synaptic connectivity and (b) KMS matrix $[\xx^{\bullet|\b}]$ for $4.38 \le\b \le 4.6$. Node size reflects incoming connections. Locomotor neurons, despite few incoming synapses, dominate incoming flow at high temperature.}
 	\label{fig:locomotors}
 \end{figure}

 
 Strikingly, within $4.38 \le\b \le 4.6$, locomotor neurons~\cite{chalfie1985neural} 
 emerge as the primary hubs of incoming flow (Fig.~\ref{fig:locomotors}b), despite their low synaptic connectivity (Fig.~\ref{fig:locomotors}a). This 'functional hubness' reveals a critical layer of organization beyond physiological properties and synaptic weights, providing the first theoretical proof of locomotion's central role in \cels' function~\cite{Piggott2011,Bono2005} and raising fundamental questions about how path-structured flow shapes brain's function.
 
\subsubsection{Flow selectivity}
\begin{figure}[!h]
	\includegraphics[width=.45\textwidth]{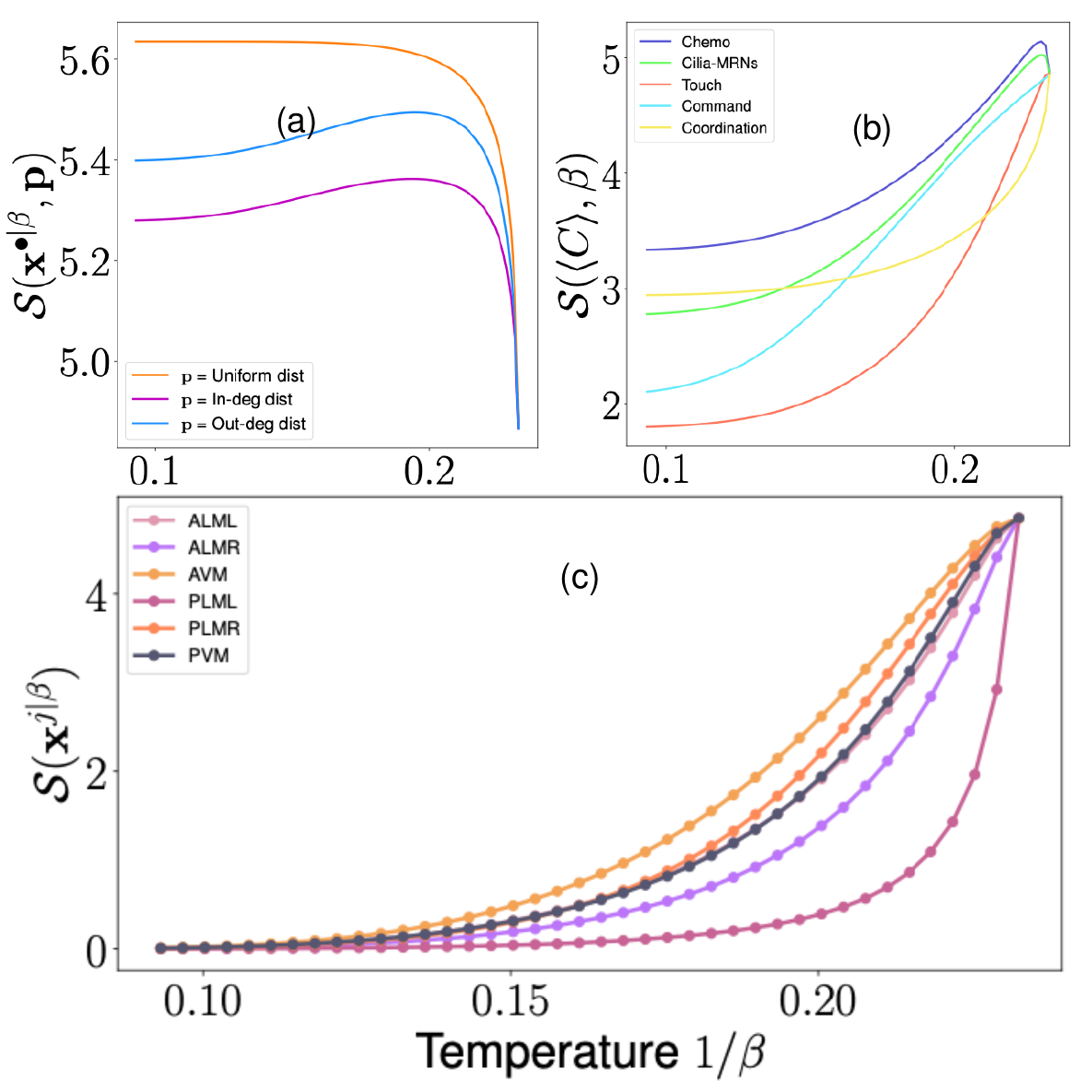}
	\caption{Equilibrium entropy. (a) $\cal S(\xx^{\bullet| \b}, \pp)$ vs. temperature $1/\b$ for different distributions: uniform (orange), in-degree (blue), and out-degree (purple). (b) Entropy of mixed states from uniform distributions over neuron subsets. (c) Entropy of individual touch-receptor neurons.}
	\label{fig:entropy}
\end{figure}

In~\ref{fig:entropy}b, six neuron subsets are compared, with touch receptors showing the highest interaction selectivity, consistent with their sensorimotor role~\cite{Goodman2006,patil2015neural}. Coordination neurons (VD and DD) also display selective outgoing path-structured flow, reflecting their role in mediating locomotion~\cite{chalfie1985neural,Bono2005}. Within the touch subset (Fig.~\ref{fig:entropy}c), PLML is the most selectivity, maintaining low entropy up to the critical temperature.

\subsubsection{Symmetry} 
\begin{figure}[!h]
	\includegraphics[width=.45\textwidth]{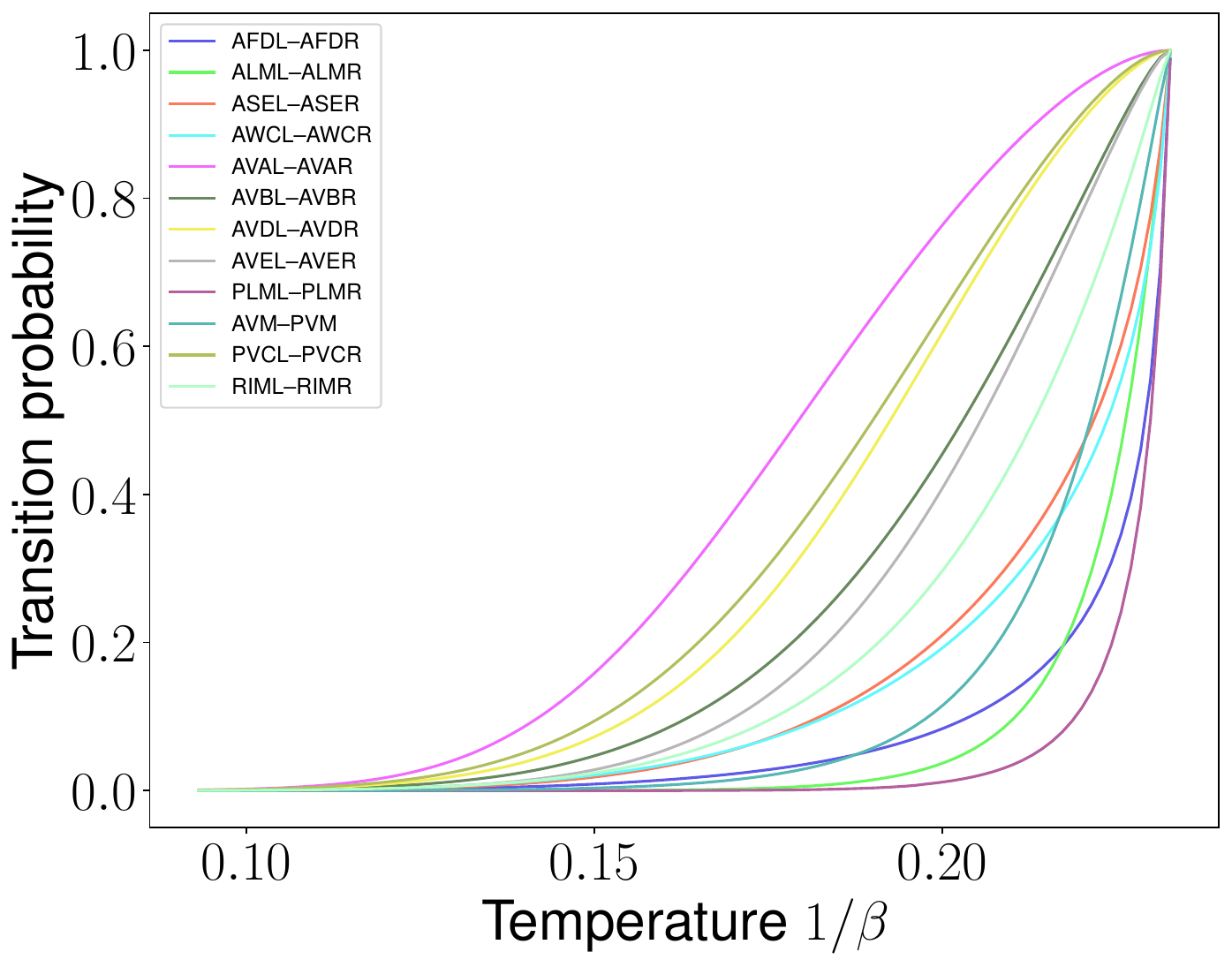}
	\caption{Transition probability $P(j,j',\b)$ between KMS states fo selected neuron pairs. Near $T_c=1/\b_c$, command interneurons (AVA, AVB, AVD, PVC) show strong left-right symmetry. Sensory neurons break symmetry below $T_c$, with PLM showing the greatest asymmetry.}
	\label{fig:transition}
\end{figure}

Fig.~\ref{fig:transition} shows left/right (or anterior/posterior) symmetry in selected neuron classes. The fact that $P$ tends to $1$ at $T_c$ for all the neuron pairs  illustrates the critical behavior of the equilibrium at this temperature; \ie, all KMS states converge to a common state at $T_c$.

We observe a high level of left-right symmetry in command interneuron classes. Whereas, the posterior touch sensory neurons PLML and PLMR have highly asymmetric  outgoing neuronal interactions, in agreement with the fact that PLML is more selective than its right counterpart (cf. Fig.~\ref{fig:entropy}c). Moreover, that $P$ tends to $1$ at $T_c$ for all the neuron pairs  illustrates the critical behavior of the equilibrium at this temperature--\ie, all KMS states converge to the unique state at $T_c$.

\subsection{Conclusion and discussion}  
Over the past three decades, graph algebras have been extensively developed in operator algebras and non-commutative geometry~\cite{cuntz1980,Pask2006}. However, their potential in interdisciplinary fields such as complex networks and neuroscience has remained largely unexplored due to their abstract nature. In this work, we showed that KMS states on graph C*-algebras provide a rigorous framework for modeling path-structured flow and analyzing their dynamics. Using empirical data from \cels, we demonstrated how KMS states generate probability distributions that capture neuronal interactions and their modulation by concentration parameters. This approach is novel in bridging algebraic methods from quantum statistical physics~\cite{hugenholtz1972,bratteli1982equilibrium}, graph C*-algebras~\cite{raeburn2005graph}, and network (neuro)science~\cite{sporns2005}, while also introducing a natural entropy measure for quantifying selectivity in flow and linking offering structure to function.

Notably, the centrality of the locomotion circuitry in Fig.~\ref{fig:locomotors} emerges at $4.38\le \b \le 4.6$, precisely where entropy exhibits non-trivial peaks (Fig.~\ref{fig:entropy}), suggesting that this $\b$ range may correspond to functional regimes relevant to experimental conditions. Future works will investigate how these values support the emergence of functional circuits. 

While our approach does not aim to describe the full complexity of neural dynamics, it offers algebraic and statistical tools to study the global and functional organization of directed networks. Moreover, due to the statistical nature of the links derived from KMS states (Fig.~\ref{fig:patterns}), some may not depend on the physical network's topology; a statistical analysis on random graphs could assess their significance. 
Moreover, since we considered only synaptic transmission, the profiles $\xx^{j|\b}$ neglect the contribution of extrasynaptic signaling to functional connectivity in \cels~\cite{ripoll2022,randi2023celegans}. These questions are addressed in our companion work which focuses on the functional implications of the framework~\cite{Moutuou2024b}. Extending our framework to multiplex networks~\cite{de2013mathematical,moutuou2023} using, for instance, the higher-rank graph algebras formalism~\cite{Raeburn2003,Christensen2020} 
could provide an integrated representation of neuronal communications.

\subsection*{Code availability} 
The code developed for this work is publicly available at \href{https://github.com/elkMm/KMSnet}{https://github.com/elkMm/KMSnet}.

\appendix{

\input{SM/SM-kms-network}

}

\begin{acknowledgments}
This work was supported by the Natural Sciences and Engineering Research Council of Canada through the CRC grant NC0981.
\end{acknowledgments}

%


\end{document}

%% file: SM/figures/toy.tex
\begin{tikzpicture}[scale=.3]
\node[vertex] (v1) at (4,0) {};
\node at (4, 0) {$1$};
\node[vertex] (v2) at (0,-3) {};
\node at (0, -3) {$2$};
\node[vertex] (v3) at (3,4) {};
\node at (3, 4) {$3$};
\node[vertex] (v4) at (-7,0) {};
\node at (-7, 0) {$4$};
\node[vertex] (v5) at (-3,1) {};
\node at (-3, 1) {$5$};
\node[vertex] (v6) at (-3,8) {};
\node at (-3, 8) {$6$};

\draw[edgeto] (v1) to [bend left=60]  (v2);
\draw[edgeto] (v1) to [bend left=40]  (v2);
\draw[edgeto] (v1) to [bend right=20]   (v3);
\draw[edgeto] (v1) to [bend left=20]  (v5);
\draw[edgeto] (v2) to [bend right=30]  (v3);
\draw[edgeto] (v2) to [bend left=50]  (v5);
\draw[edgeto] (v5) to [bend right=10]  (v2);
\draw[edgeto] (v2) to [bend left=30]  (v4);
\draw[edgeto] (v3) to [bend right=10]   (v4);
\draw[edgeto] (v5) to [bend right=5]   (v3);
\draw[edgeto] (v5) to [bend right=10]  (v4);
\draw[edgeto] (v5) to [bend left=30]   (v6);
\draw[edgeto] (v3) to [bend right=-5] (v6);
\draw[edgeto] (v3) to [bend right=40]  (v6);
\draw[edgeto] (v3) to [bend right=90] (v6);
\draw[edgeto] (v4) to [bend left=30] (v6);
\draw[edgeto] (v2) edge  [loop below]  (v2);
\draw[edgeto] (v4) edge  [loop below]  (v4);
\end{tikzpicture}

%% file: SM/SM-kms-network.tex
\section{Algebraic Quantum Mechanics}\label{Appendix:AQM}	
We present here an overview about Algebraic Quantum Mechanics (AQM) and the Toeplitz-Cuntz-Krieger (TCK) algebra of directed networks~\cite{bratteli1982equilibrium,Raeburn2003} and provide details about our KMS states formalism for quantifying interaction properties in directed networks.	
\subsection{Basics of C*-algebras}
 A $C^*$--{\em algebra} is Banach algebra $\cal O$ over $\bb C$ equipped with an involution 
\[
\cal O \rTo \cal O, \ A \mto A^*, 
\]
such that
\begin{itemize}
	\item[(i)] $\|UV\| \le \|U\|\|V\|$ for all $U, V\in \cal O$, and 
	\item[(ii)] $\|U^*U\|=\|U\|^2$ for all $U\in \cal O$.
\end{itemize}
We say $\cal O$ is {\em unital} if it has a unit element $1$. An element $U\in \cal O$ is {\em self-adjoint} if $U^*=U$. And $U$ is said to be {\em positive} if $U=V^*V$ for some $V\in \cal O$. In such a case, one writes $U\ge 0$. Moreover, we write $U\ge V$ if $U-V$ is positive in $\cal O$.

For instance the matrix algebra $\cal O = M_n(\bb C)$ is a unitial $C^*$-algebra where the involution is given by the matrix transpose $A^*=A^\top$, for $A\in M_n(\bb C)$, and $1$ is the identity matrix. More generally, the algebra $B(\cal H)$ of bounded linear operators on a Hilbert space $\cal H$ is a $C^*$--algebra with respect to the operator norm $\|\cdot\|_{op}$ given by
\[
\|T\|_{op} = \sup\{\|T\xi\|: \xi\in \cal H, \ \|\xi\|\le 1\}.
\]
An operator $T\in B(\cal H)$ is positive if and only if $\inner{T\xi}{\xi} \ge 0$ for all $\xi\in \cal H$.

\subsubsection{States}
 A linear functional $\o:\cal O\rTo \bb C$ is said to be {\em positive} if its values on positive elements are all positive; that is,  $\o(U^*U)\ge 0$ for all $U\in \cal O$. A {\em state} on a $C^*$-algebra $\cal O$ is a positive linear functional $\o$ such that $\|\o\|=\sup_U|\o(U)| = 1$.  A {\em trace-state} on $\cal O$ is a state $\o$ satisfying $\o(UV)=\o(VU)$.

For example, the usual trace of a matrix is a trace-state on $M_n(\bb C)$, showing that the notion of state is, in fact, a generalization of the trace functional.

\subsubsection{C*-dynamical systems}
 The mathematical formalism of quantum statistical mechanics in the realm of $C^*$-algebras~\cite{bratteli1982equilibrium, hugenholtz1972, jaksic2002} and noncommutative geometry~\cite{connes1995ncg} can be summarized as follows. A quantum system is represented by a pair $(\cal O, \a)$ --called a $C^*$--{\em dynamical system}--  consisting of a $C^*$-algebra $\cal O$ and the time evolution $(\a_t)_{t\in \bb R}$ of the system, which is a one-parameter group of $*$-automorphisms of $\cal O$; that is, $\a$ is a continuous map from $\bb R$ to the automorphism group of $\cal O$ that respects the C*-algebra structure~\cite{dixmier1982c}. Self-adjoint elements of $\cal O$ usually referred to as the {\em observables} of the system.
 
  In the finite dimensional case where $\cal O$ is a matrix algebra $M_n(\bb C)$, the one-parameter group of $*$-automorphisms in a $C^*$--dynamical system $(M_n(\bb C), \a)$ is an action of the form 
\begin{eqnarray}
	\a_t(A) = e^{itH}Ae^{-itH}, \ t\in \bb R, A\in M_n(\bb C),
\end{eqnarray}
where $H\in M_n(\bb C)$ is a self-adjoint matrix, and we recover the usual Hamiltonian of a quantum system~\cite{Hall2013}.

Now given a state $\o$ on $\cal O$, one might think of the element $\o(U)$ as the expectation value of the observable $U$ when the system is in state $\o$. By analogy, the physical interpretation of $\o(\a_t(\cdot))$ is that the system is in state $\o\circ \a_t$ at time $t$ if it was in state $\o$ at time $0$.

\subsubsection{KMS states}
 The thermal equilibrium states of the system at inverse temperature $\b=\frac{1}{T}$ are mathematically characterized by the {\em Kubo-Martin-Schwinger (KMS)} conditions~\cite{haag1967quantum, woronowicz1985KMS}. Specifically, a state $\o$ on $\cal O$ is a {\em KMS state} at inverse temperature $\b$ on $(\cal O, \a)$, or a $\b$--{\em KMS state} in short, iff for all $U, V\in \cal O$, there is an analytic function $F_{U,V}(z)$, bounded and continuous on the strip $0<Im(z)<\b$, such that 
\begin{eqnarray}\label{eq:KMS-equations}
\begin{aligned}
F_{U,V}(t) & =  \o(U\a_t(V)), \ \forall t\in \bb R \\
 F_{U,V}(t+i\b)& =  \o(\a_t(V)U), \ \forall t\in \bb R.
 \end{aligned}
\end{eqnarray}
These conditions are equivalent to $\o$ satisfying
\begin{eqnarray}\label{eq:KMS-condition}
\o(U\a_{i\b}(V))= \o(VU),
\end{eqnarray}
for all analytic elements $U,V$ in $(\cal O, \a)$, where an element $U\in \cal O$ is said to be {\em analytic} in the $C^*$-dynamical system if the function $t\mto \a_t(U)$ extends to an analytic function 
\[
\bb C \rTo \cal O, \ t+i\b\mto \a_{t+i\b}(U).
\]
It is immediate that a $\b$-KMS state $\o$ is necessarily time-invariant with respect to the dynamics $\a$; that is, $\o(\a_t(U))=\o(U)$ for all $U\in \cal O$.
 In particular, $0$-KMS states are {\em trace-states} (see Supplementary Note 1) that are time-invariant, and they represent the equilibrium states of the system at infinite temperature. And at the other extreme, $\infty$-KMS states correspond to the {\em ground states} of the systems~\cite{bratteli1982equilibrium, haag1967quantum}.

\subsubsection{Relation to Gibbs states}
In the finite-dimensional case of a matrix algebra where the dynamics is defined by a Hamiltonian $H$, for any non-negative inverse temperature $\b$, the $C^*$-dynamical system $(M_n(\bb C), \a)$ has a unique $\b$-KMS equilibrium state given by the Gibbs state~\cite{Hall2013}
\begin{eqnarray}
	\o_{\b}(A) = \frac{\Tr(e^{-\b H}A)}{\Tr(e^{-\b H})},
\end{eqnarray}
for $A\in M_n(\bb C)$.

\subsection{Directed networks as algebraic quantum systems}

\subsubsection{Directed networks}
A directed complex network is represented by a directed graph $G$ consisting of a finite set of nodes $V=\{i, j,...\}$, and a finite set of directed edges $E = \{a, b, ...\}$ between pair of nodes. If $a: i\lto j$ is an edge from $j$ to $i$, we write $\pre(a)=j$ and $\post(a)=i$. We allow $G$ to be a  '{\em multigraph}'; that is, there might be more than one edge from a node to another (parallel edges), and it is possible to have edges from one node onto itself (self-loops). \Cref{fig:toy} shows a toy example of such graphs. A {\em walk} of length $n$ is a sequence $\g = a_n\cdots a_1$ of edges $a_1, ...,a_n$ such that $\pre(a_{l+1}) = \post(a_{l})$ for $l=1, ...,n-1$. For a walk $\g$, we write $|\g|=n$, $\pre(\g)=\pre(a_1)$, and $\post(\g) = \post(a_n)$. We let $N=\# V$ be the number of nodes of $G$, and define the {\em adjacency matrix} $A\in M_N(\bb N)$ of $G$ by
\begin{eqnarray}
	A_{ij} := \#\{a: \pre(a)=j, \ \post(a)=i\}.
\end{eqnarray}
Observe that for a non-negative integer $n$, $A_{ij}^n$ is the number of directed walks of length $n$ from $j$ to $i$.

\subsubsection{The TCK algebra}
 An efficient and elegant way to represent and study the combinatorial and topological properties of a directed network is to view its nodes and edges as {\em projections} and {\em partial isometries} on an infinite dimensional Hilbert space $\cal H$ to generate a C*-algebra encoding all possible routes within the graph. 
 
 We first recall that a {\em projection} on the Hilbert space $\cal H$ is a self-adjoint bounded linear operator $Q$ such that $Q^2=Q$, and a {\em partial isometry} is a bounded linear operator $S$ on $\cal H$ such that $S^*S$ is a projection (which is equivalent to requiring $SS^*$ to be a projection)~\cite{dixmier1982c}. For instance, let $\cal H$ be the Hilbert space of all complex-valued sequences $h=(\xi_\g)_\g$ indexed on walks of finite lengths in $G$ such that $\|h\|^2 = \sum_\g|\xi_\g|^2 < \infty$. Note that the scalar product $\< , \>$ of $\cal H$ is given by $\< \xi, \xi'\> = \sum_\g \xi_\g \bar{\xi'}_\g$ for $h=(\xi_\g)$ and $h'=(\xi_\g')$. Now, for each node $i$ in $G$, define the map $Q_i: \cal H\rTo \cal H$  such that for $h=(\xi_\g)_\g \in \cal H$, 
 \begin{eqnarray}
 	Q_i(h) = (\xi_\g)_{\post(\g)=i}.
 \end{eqnarray}
The intuition here is that, given a (possibly infinite) sequence of (re)scaled walks, $Q_i$ is the operation that consists of selecting only the scaled walks that end at node $i$.  Namely, $Q_i$ is the projection of $\cal H$ onto the subspace $\cal H_i$ of all sequences indexed over all walks ending at node $i$. Indeed, it is immediate that $Q_i^2 = Q_i$ and that $$\<Q_ih,h'\> = \sum_{\post(\g)=i}\xi_\g\bar{\xi}_\g=\<h,Q_i^*h'\>,$$ hence $Q_i^*=Q_i$. Observe that these projections are mutually orthogonal; that is, $Q_iQ_j=0$ for $i\neq j$. 
 
On the other hand, for each edge $a:i \lto j$ in $G$, let $S_a$ be the bounded linear operator defined on $\cal H$ by 
 \begin{eqnarray}
 	S_a(h) = S_a\left((\xi_\g)_\g\right) = \left(\xi_{a\g}\right)_{\post(\g)=j},
 \end{eqnarray}
 for a sequence $h=(\xi_\g)_\g\in \cal H$. That is, $S_a$ is the operation that first select all scaled walks ending at $j$ and then extend them to walks ending at $i$ scaled at the same factors. Using the inner product of $\cal H$, we then get
 \[
 \begin{aligned}
 	\<S_a(h),h'\> = & \<(\xi_{a\g})_{\post(\g)=j}, (\xi'_{\g'})_{\g'}\> \\
 	 = & \sum_{\post(\g)=j}\xi_{a\g}\bar{\xi}'_{a\g} \\
 	 = & \<h, S_a^*(h')\>,
 \end{aligned}
 \]
therefore, the adjoint operator $S_a^*$ is given by 
\[
S_a^*(h') = \left(\xi'_{\g}\right)_{\post(\g)=i, \g = a\g'}.
\] 
This means that $S_a^*$ is the operation that select all walks that end at $i$ passing through $a$ and 'reduce' them into walks ending at $j$. Now one can check that 
\begin{eqnarray}\label{eq:TCK}
	\begin{aligned}
		S_a^*S_a & = Q_j, \ {\rm for \ all \ } a: i\lto j,\\
		\sum_{\post(a)=i}S_aS_a^* & \le Q_i , \ {\rm for \ all \ node \ } i.
	\end{aligned}
\end{eqnarray}
In particular, the operators $S_a$ are partial isometries. The relations~\eqref{eq:TCK} are referred to as the {\em TCK conditions} in the literature~\cite{kumjian1998, raeburn2005graph}. The {\em TCK algebra} (or the graph algebra) of $G$ is the {\em universal} C*-algebra $\cal O_G$ generated by the family $\{Q_i, S_a\}$ and the TCK conditions. Wherever the graph $G$ is understood, we will drop the subscript and just write $\cal O$. The universal property here means the construction of $\OG$ does not depend on the specific choice of the Hilbert space $\cal H$ neither the particular definitions of the projections $Q_i$ and partial isometries $S_a$. More precisely, any family $\{Q_i, S_a\}$ of mutually orthogonal projections $Q_i$ and partial isometries $S_a$ satisfying the TCK conditions generate a C*-algebra that is isomorphic to $\OG$~\cite{kumjian1998}. 

Note that, by definition, $\OG$ is the completion of the space spanned by projections of the form $S_{\g}S_{\t}^*$ such that $\pre(\g)=\pre(\t)$, where for a walk $\g = a_n\cdots a_1$, we set $S_\g:= S_{a_n}\cdots S_{a_1}$. Moreover, denoting $Q = \sum_iQ_i$, we have for $S_\g S_\t^*\in \OG$ and $h=(\xi_\mu)_\mu\in \cal H$ ,
\begin{eqnarray}\label{eq:identity}
	\begin{aligned}
	Q(S_\g S_\t^*)(h) = & Q\left((\xi_{\g\g'})_{\g''=\t\g', \post(\g')=\pre(\g)}\right) \\
		= & Q_i\left((\xi_{\g\g'})_{\g''=\t\g', \post(\g')=\pre(\g), i = \post(\g)}\right) \\
		= & \left((\xi_{\g\g'})_{\g''=\t\g', \post(\g')=\pre(\g)}\right) \\
		= & S_\g S_\t^*(h).
	\end{aligned}
\end{eqnarray}
With similar calculations, one can show that $(S_\g S_\t^*)Q(h)  = (S_\g S_\t^*)(h)$,  which implies that $\OG$ is a unital C*-algebra with unit element $Q$.

\subsubsection{The $C^*$--dynamics}
 There is a natural one-parameter group of *--automorphisms $(\a_t)_{t\in \bb R}$ on $\OG$ given by
\begin{eqnarray}
	\begin{aligned}
		\a_t(Q_i) & = Q_i \ i \in V, \\
		\a_t(S_a) & = e^{it}S_a, \ a\in E, 
	\end{aligned}
\end{eqnarray}
thus defining an algebraic  quantum system represented by the C*--dynamical system $(\OG, \a)$.

\subsubsection{KMS states on the TCK algebra}
 From the KMS condition~\eqref{eq:KMS-condition}, a state $\o$ of $\OG$ is KMS at inverse temperature $\b>0$ iff:
 
\begin{eqnarray}\label{eq:KMS-graph}
	\o(S_{\g}S_{\t}^*) = \d_{\g, \t}e^{-\b |\g|}\o(Q_{\pre(\g)}), 
\end{eqnarray}
for all walks $\g$ and $\t$ with $\pre(\g)=\pre(\t)$. In particular, for an edge $a: i\lto j$, we get
\begin{eqnarray}\label{eq:KMS-graph2}
	\o(S_aS_a^*)=e^{\b}\o(Q_j).
\end{eqnarray}

It follows from~\ref{eq:TCK} and~\ref{eq:KMS-graph2} that if $\o$ is a $\b$-KMS state on $\OG$, then for all node $i$, we have $\sum_{\post(a)=i}e^{-\b}\o(Q_{\pre(a)}) \le \o(Q_i) $; in other words,
\begin{eqnarray}\label{eq:state-ineq}
		\sum_{\post(a)=i}\s(Q_{\pre(a)}) \le & e^{\b}\o(Q_i).
\end{eqnarray}
Moreover, by observing that 
\[
\begin{aligned}
	\sum_{\post(a)=i}\o(Q_{\pre(a)}) & = \sum_{j}\sum_{a: i\lto j}\o(Q_j)\\
	& =\sum_jA_{ij}\o(Q_j),
\end{aligned} 
\]
and defining the vector $\xx^{\o}\in \bb R_+^N$ by the non-negative numbers $\xx^{\o}_i := \o(Q_i)$, the inequality~\eqref{eq:state-ineq} yields

\begin{eqnarray}\label{eq:subeigenvector}
	A\xx^{\o} \le e^{\b}\xx^{\o},
\end{eqnarray}
for a $\b$-KMS state $\o$, where the inequality is coordinate-wise. One might then think of $\xx^{\s}$ as a '{\em sub-eigenvector}' of $A$ for $e^\b$.

Now since the unit element of $\OG$ is given by $\sum_iQ_i$, we obtain $\o(\sum_iQ_i)=1$. Hence,
\begin{eqnarray}
	\|\xx^{\o}\|_1:= \sum_i\xx^{\o}_i = 1,
\end{eqnarray}
which implies that $\xx^{\o}$ is a probability distribution on the nodes set $V$.


\subsubsection{The partition vector}
 Le $b_c = \log r$, where  $r$ be the spectral radius of $A$; \ie 
 \[
 r = \max \{|\l|, \ \l \ {\rm eigeinvalue \ of \ } A\}.
 \] 
 As a consequence of Equation~\eqref{eq:subeigenvector}, if $\b> \b_c$ and $\o$ is a $\b$-KMS state, then $e^{\b}$ is not an eigenvalue of $A$ and $\xx^\o$ is not an eigenvector of $A$. Hence, the inequality becomes strict. We call $\b_c$ the {\em critical inverse temperature} of $G$. It follows that for $\b> \b_c$ the vector 
\begin{eqnarray}\label{eq:psi}
	\PSI^\o := (1-e^{-\b}A)\xx^{\o}
\end{eqnarray}
is strictly positive, and
\begin{eqnarray}\label{eq:xi-psi}
	\xx^{\o}=(1-e^{-\b}A)^{-1}\PSI^{\o}
\end{eqnarray}
Moreover, since $e^{-\b}r<1$, it follows from the theory of functional analysis (see for instance~\cite[Chap. 1 \& 5]{buhler2018functional} that 
\begin{eqnarray}\label{eq:resolvent}
	(1-e^{-\b}A)^{-1} = \sum_{n=0}^\infty e^{-\b n}A^n.
\end{eqnarray}
Hence, $\xx^{\o} = \sum_ne^{-\b n}A^n\PSI^{\o}$. Now consider the coordinate-wise non-negative vector $Z^\b$ indexed over the nodes set $V$ and given by
\[
Z^\b_j = \sum_{\pre(\g)=j}e^{-\b |\g|}.
\]
To see that these quantities are well defined, observe that the right-hand side can be written as
\[
\begin{aligned}
	\sum_{\pre(\g)=j}e^{-\b |\g|} & = \sum_b\sum_{|\g |=n, \pre(\g)=j}e^{-\b n} \\
	& = \sum_n\sum_{i}e^{-\b n}A^n_{ij} \\
	& = \sum_{i}\sum_n e^{-\b n}A_{ij}^b,
\end{aligned}
\]
and, thanks to~\eqref{eq:resolvent}, this infinite series converges to $$ \sum_i (1-e^{-\b}A)^{-1}_{ij}$$ when $\b > \b_c$. Therefore the {\em partition vector}  $Z^\b$ can be expressed in terms of the adjacency matrix by
\begin{eqnarray}\label{eq:emittance2}
	Z^\b_j = \sum_i(1-e^{-\b}A)^{-1}_{ij}.
\end{eqnarray}
The number $Z^\b_j$ represents the total volume of all flow pathways coming from node $i$ at inverse temperature $\b$, given that the functioning of each link in the network is affected by a factor of $e^{-\b}$.  

\subsubsection{From the TCK algebra to the underlying network} 
Notice that $\<\PSI^{\o}, Z^{\b}\>=\|\xx^\o\|_1=1$. Indeed, using~\eqref{eq:resolvent}, we have
\[
\begin{aligned}
	\sum_i\xx_i^\o & = \sum_i\sum_n \sum_j e^{-\b n}A^n_{ij}\PSI_j^\o \\
	& = \sum_j \PSI_j^\o \left[ \sum_i\sum_n e^{-\b n} A^n_{ij} \right] \\
	& = \sum_j \PSI_j^\o\left[ \sum_{\pre(\g) = j}\sum_n e^{-\b |\g |} \right] \\
	& = \sum_i\PSI_i^\o Z^\b_i \\
	& = \<\PSI^\o, Z^\b\> \\
	& = 1.
\end{aligned}
\]
This proves that the distribution $\xx^\o$ associated with the $\b$--KMS state $\o$ is completely determined,  through~\eqref{eq:xi-psi}, by the vector $\PSI^\o$ that is a solution of the following equation.
\begin{eqnarray}\label{eq:psi-emittance}
	\<\PSI, Z^{\b}\> = 1.
\end{eqnarray}

Conversely, it was proved by an Haef {\em et al.}~\cite{huef2012KMS} that for any non-negative vector $\PSI\in \bb R_+^N$ satisfying~\eqref{eq:psi-emittance}, with $\b > \b_c$, there is a $\b$-KMS state $\o=\o_{\PSI}$ given by 
\[
\o(S_{\g}S_{\t}^*) = \d_{\g, \t}e^{-\b |\g|}\xx_{\pre(\g)},
\] 
where $\xx\in \bb R_+^N$ is given by $\xx = (1-e^{-\b}A)^{-1}\PSI$ as in formula~\eqref{eq:xi-psi}. Furthermore, this process defines a one-to-one correspondence between the set of such vectors $\PSI$ and the $\b$-KMS states on $\OG$.

\subsubsection{KMS states matrices}

 For a fixed node $j$ and $\b>\b_c$, define the vector $\PSI^{j|\b}\in [0,\infty)^N$ as
\begin{eqnarray}
	\PSI^{j|\b} = \left(\frac{1}{Z^{\b}_j}\d_{i,j}\right)_{i\in V}. 
\end{eqnarray}
It is obvious that these vectors are solutions to Eq.~\eqref{eq:psi-emittance}. Hence, we get for each fixed node $j$, a $\b$-KMS state $\xx^{j|\b}$ given by
\[
\xx^{j|\b}= \left((1-e^{-\b}A)^{-1}\PSI^{j|\b}\right)_i = \frac{1}{Z^\b_j}\sum_n e^{-\b n}A^n_{ij}.
\]
Or in a more compact form,
\begin{eqnarray}
	\xx^{j|\b} = \frac{1}{Z^\b_j}(1-e^{-\b}A)^{-1}_{ij}.
\end{eqnarray}

It can be seen that the space $\Omega_\b$ of all $\b$-KMS states on $G$ is the simplex of dimension $N-1$ generated by the vectors $\xx^{j|\b}$; that is,

\begin{eqnarray}\label{Ap:eq:KMS-simplex}
	\Omega_\b = \left\{\sum_j\pp_j\xx^{j|\b} | 0\le \pp_j\le 1, \  \sum_j \pp_j=1\right\}.
\end{eqnarray}
Thus the vectors $\xx^{j|\b}$ are the {\em pure ($\b$--KMS) states} of the system, while their convex combinations are {\em mixed states}~\cite{bratteli1982equilibrium}. In a sense, the states $\xx^{j|\b}$ are the analog of eigenvectors and the simplex $\Omega_\b$ plays the role of eigenspace. It follows that every $\b$-KMS state $\xx$ can be written as a matrix product
\begin{eqnarray}\label{eq:Zmat}
	\xx= [\xx^{\bullet|\b}]\pp,
\end{eqnarray}
where $\pp=(\pp_j)_j$ is a probability distribution on the nodes and $[\xx^{\bullet|\b}]$ is the column stochastic $N\times N$-matrix with non-negative entries whose columns are the the pure KMS states $\xx^{\bullet|\b}$; that is, $$[\xx^{\bullet|\b}]_{ij}:= \xx^{j|\b}_i.$$

\section{KMS matrices as non-Markovian processes}\label{Appendix:process}

It is straightforward to check that, for a fixed $\b > \b_c$, the KMS matrix $[\xx^{\bullet|\b}]$, formed by the pure states $\xx^{j|\b}$, naturally corresponds to a generalized stochastic process that can be represented as a {\em non-Markovian random walk with memory decay} $\{X_t\}_{t\in \bb N}$, where transitions are influenced by weighted sums over the entire history of walks, but with exponentially fading influence of older steps where the memory decay is governed by $\b$ (unlike a standard one-step Markov chain, where transitions depend only on the current node). This process is defined as follows:

\begin{itemize}
	\item a walker starts at node $j\in V$;
	\item at each time step, the walker proceeds with probability $e^{-\b}$, or stops with probability $1-e^{-\b}$;
	\item The probability of reaching node $i$ after $n$ steps is therefore weighted by $(1-e^{-\b})e^{-n \b}$ and is the same as the probability of reaching $i$ through an arbitrary path of length $n$; more precisely, we get
	\begin{eqnarray}
		\begin{aligned}
		P(X_n= i  | X_0 = j) = & \frac{ (1-e^{-\b}) e^{-n \b}A_{ij}^n}{\sum_{k\in V}\sum_{m\ge 0}(1-e^{-\b})e^{-m \b}A^m_{kj}} \\
		= & \frac{ e^{-n \b}A_{ij}^n}{\sum_{k\in V}\sum_{m\ge 0}e^{-m \b}A^m_{kj}}.
		\end{aligned}
	\end{eqnarray}
\end{itemize}

It follows that the total probability that a walker starting at $j$ {\em eventually} reaches $i$ is 

\begin{eqnarray}
	\frac{\sum_{n\ge 0} e^{-n \b}A_{ij}^n}{\sum_{k\in V}\sum_{n\ge 0}e^{-n \b}A^n_{kj}} = \xx^{j|\b}_i.
\end{eqnarray}

Therefore, under this model, the pure KMS state $\xx^{j|\b}$ is the {\em stationary distribution} over end-nodes starting from node $j$, considering all possible walk lengths with exponentially decreasing weight. In other words, these distributions measure the long-term behavior of the dynamics, capturing how information, influence, or flow propagates and accumulates in the directed network over time when paths of all lengths, redundancies and feedback  are taken into account with an exponentially decaying memory.\\

Moreover, this perspective gives a better physical intuition of the fact that high temperatures ($\b$ small) favor longer paths, since the walker is more likely to continue at each step; while low temperatures (large $\b$) will {\em localize} the dynamics by making the walker more likely to stop after one or a few steps (shorter paths).\\

By interpreting the KMS states this way, we show that our model does capture a kind of structurally embedded propagation, even though it is not time-dependent in the traditional dynamical systems sense. The “flow” arises as an emergent pattern over path ensembles, not from explicit dynamics over time.


	